\documentclass[aps,pre,onecolumn,amsmath,amssymb]{revtex4}
\usepackage{bm}
\usepackage{graphicx}

\newcommand{\be}{\begin{eqnarray}}
\newcommand{\ee}{\end{eqnarray}}

\newcommand{\pol}{{\textstyle\frac{1}{2}}}
\newcommand{\ba}{\begin{array}}
\newcommand{\ea}{\end{array}}

\newcommand{\ket}[1]{|#1\rangle}

\begin{document}

\title{Superfast Algorithms and the Halting Problem in Geometric Algebra}
\author{Marcin Paw\l owski}

\address{Katedra Fizyki Teoretycznej i Informatyki Kwantowej\\
Politechnika Gda\'nska, 80-952 Gda\'nsk, Poland}

\begin{abstract}
A new type of algorithms is presented that combine the advantages of quantum and classical ones. Those combined advantages
along with aspects of Geometric Algebra that open possibilities unavailable to both of these computations
are exploited to obtain database search and number factoring algorithms that are faster than the quantum ones, and even to
create a ''pseudoalgorithm'' that can perform noncomputational tasks.
\end{abstract}

\maketitle

\section{Introduction}

Recently it has been shown in \cite{AC} that quantum algorithms can be translated into the structures of Geometric Algebra (GA). An example
of Deutsch-Jozsa algorithm has been presented there along with a remark that all quantum algorithms are expected to have GA analogues.
But GA has some properties not encountered in the quantum world. For example, any information about the state can be easily
extracted at any time. In the case of quantum computation the lack of the ability to do that is the main source of the
complexity. That difference is exploited in the next two sections to create database search and number factoring algorithms
that are much faster than their quantum versions. Then another remarkable property of GA is being used. Namely, the fact that
multivectors can be $\infty$-dimensional will allow a construction of a pair of multivectors: one
being a coded Universal Turing Machine and its input, the other a ''pseudoalgorithm''. The analysis of their geometric product gives
the answer to the Halting Problem which is known to have no algorithmic solution.
\

No explanation of algebraic properties of Geometric Algebra or Clifford Algebras is given here.
For a brief but sufficient introduction reader should turn to \cite{GA},\cite{AC} or \cite{ACM}.
Also some issues concerning Turing Machines are omitted, for example: how a Turing Machine is coded into the number $m$ or what
$U_i$ does when it is ordered to operate on meaningless data. There are multiple solutions to these problems, but since
the choice made does not influence any conclusions drawn in this paper we will not chose any specific one.
\

Finally, it is important to understand that the aim of this paper is not to give any example of physical systems that
are able to perform discussed tasks, nor to convince the readers that such systems are bound to exist. The aim is to
show the power of Geometric Algebra and what could be in principle done with systems described by it.

\section{Database search algorithm}
We assume that there exists an oracle that can recognize a certain group of states $x\in X$. The action of this oracle will be given
by:
\be
O_f\ket{x}\ket{y}=\ket{x}\ket{y\oplus f(x)}
\ee
The function $f(x)$ is defined as being 1 if $x\in X$ , and 0 if $x\notin X$. Our task is to
find the states from X in a given set Y. Classical computers require O(N) operations to perform this task. Quantum ones need
only O($\sqrt{N}$) \cite{NC}, but using GA we are able to solve this problem in just two steps regardless
of the database size.
\

First we will need a method of coding numbers into multivectors. It is very convenient to use the method proposed in
\cite{ACM}: $n$-bit number $x$ which $i$-th digit of the binary form is $x_i$ is coded as a k-blade, that is the geometric product
of $k$ basis vectors, where $k$ is the number of 1's in the binary representation of $x$. The explicit form of the coded number is:
\be
L(x)=\prod_{i=1}^n (e_i)^{x_i}
\ee

The algorithm will operate in $n+1$ dimensional space, where $n$ dimensions will be used to store the states, that is the coded numbers
$L(x)$, and the last dimension will contain the analogue of the oracle qubit. Since there is no tensor product in GA, the action of
the oracle has to be defined in another way. The most natural way would be:
\be
O_f L(x)(e_{n+1})^y=L(x)(e_{n+1})^{y\oplus f(x)}
\ee

The initial state is:
\be
\psi_0=\sum_{x\in Y} L(x) (1-e_{n+1})
\ee

It is easy to see that:
\be
O_f L(x)(1-e_{n+1})&=&-L(x)(1-e_{n+1}) \quad \quad  \textrm{if} \quad x\in X \quad \textrm{and}
\\
O_f L(x)(1-e_{n+1})&=&L(x)(1-e_{n+1}) \quad \quad  \ \  \textrm{if} \quad x\notin X
\ee

The operation that retrieves the needed states from $\psi_0$ is:
\be
\pol(1-O_f)\psi_0=\sum_{x\in X} L(x) (1-e_{n+1})
\ee

Now it is enough to look at the subspace spanned by the first $n$ basis vectors to get the states that were searched for.
\

It is easy to show that using parallelism and transparency (the explanation of this terms is given in the summary)
of GA a wide range of problems can be solved much faster than
it is currently possible. That is done in the following section using the example of number factoring algorithm.

\section{Number factoring}

First we will show that any classical algorithm can be translated into the GA language. To do this it is enough to present
a way of coding the NAND gate which is known to be universal for classical computation.
\

Let's divide the space that the algorithm will operate in into $N$ $n$-dimensional subspaces. The $m$-th subspace is spanned
by vectors $\{e_{mn+1},....,e_{(m+1)n}\}$. Let $Lm(x)$ denote $n$ digit number $x$ coded as a blade from the subspace $m$. If coding
$2n$ digit number is necessary it will require two subspaces and will be denoted as $Lm,p(x)$.
The whole space can be considered as a $Nn$ bit memory that contains 1 at $k$-th place if $e_k$ appears in the geometric product
that is the blade and 0 otherwise. To perform any calculation possible for classical computation it suffices to construct
an operation that puts in the empty (containing 0) place $r$ the NAND of digits from places $p$ and $q$ and leaves the rest of the
space intact.
\

Let $P_k(M)=P_{e_k}(M) $ where the RHS is defined as in \cite{LD}. Then $e_k\cdot P_k(M)$ is equal to the number stored
in the $k$-th place of the memory. It is easy to see that operation:
\be
NAND(p,q,r,M)=\Big((e_p\cdot P_p(M))(e_q\cdot P_q(M))+e_r\big(1-(e_p\cdot P_p(M))(e_q\cdot P_q(M))\big)\Big)M
\ee
acts on the memory $M$ blade putting in $r$-th place the NAND of the digits from places $p$ and $q$. This proves that every classical
computation can be translated into the GA language. For example, we can construct operation $T$ that multiplies the numbers:
\be
T(L0(x)L1(y)L2(y))=L0,1(xy)L2(y)
\ee
To expand the powers of computation the linear operator $T'$ is needed that will perform the operation of $T$ on every blade
of the multivector separately and sum the outputs up. It is possible to find such $T'$ for every $T$
\be
T\sum_x\sum_y L0(x)L1(y)L2(y)=\sum_x\sum_y L0,1(xy)L2(y)
\ee
Projecting the RHS of this equation onto $L0,1(Z)I2$ where $I2$ is the $n$-blade from subspace 2 we get:
\be
\sum_y L0,1(Z)L2(y)
\ee
Looking at subspace 2 we see every divisor of $Z$. This algorithm factorizes $Z$ with the speed that is bounded by the time
required to multiply two $n$-digit numbers. This is an outstanding result but it is not the limit of
GA algorithms powers.

\section{The Halting Problem}

It is known that there is no algorithm to solve the Halting Problem that is to answer whether or not given Turing Machine
will ever stop during the calculation with given input data. The operator that solves this problem is given below, but
there is no contradiction with the preceding statement since this operator is not an algorithm, since its simulation on
any classical or quantum computer would require infinite time to complete.
\

In the previous section it was shown that any classical computation can be done in GA. So it is possible to construct
an operation $U_i$ that acts in the $\infty$-dimensional space, that is divided into infinite number of subspaces
(each of them also of infinite dimension), as an Universal Turing Machine. In the subspace 0 number $m$ is coded which is
the number corresponding to some Turing Machine. The operation $U_i$ takes the machines memory $x$ from the subspace
$4i+1$ and the machines state $s$ form subspace $4i+2$ and calculates one step of the Turing Machine $m$ with memory $x$ and state $s$
putting new state of the memory in subspace $4i+3$ and new state of the machine in subspace $4i+4$.
\

Now lets consider operator
\be
U=\prod_{i=0}^\infty U_i
\ee
acting on the initial state
\be
\psi_0=\Big(\sum_{m=0}^\infty L0(m)\Big)\prod_{i=0}^\infty\Big(\sum_{x=0}^\infty L(4i+1)(x) \sum_{s=0}^\infty L(4i+2)(s) \Big)
\ee
Its action can be viewed as calculating a single step of every Turing Machine in every possible (or not) state, and every possible (or not)
state of memory and writing it down infinite number of times.
\

The state
\be
\psi_1=U\psi_0
\ee
is a sum of blades. Each codes the number of a certain Turing Machine and a sequence of machine and memory states. To choose
the sequences that actually make sense a projection $P$ is required. $P$ projects into the subspace spanned by all blades for which
for every $i$ $L(4i+3)(x)=L(4i+7)(x)$ and $L(4i+4)(x)=L(4i+8)(x)$.
\

The state
\be
\psi_2=P\psi_1
\ee
fully describes the action of every Turing Machine with every possible initial state of memory and machine. To
solve the Halting Problem for any particular Turing Machine $m'$ with particular input $x'$ we first act with the
projector $P(m',x')$ which projects the blades into the subspace spanned by blades which have $m'$ coded into 0-th subspace,
$x'$ into first and number describing the initial state into second.
\

The state
\be
\psi_3=P(m',x')\psi_2
\ee
has the state of $m'$-th Turing Machine after $i$ steps coded in subspace $4i+2$. Now it is enough to act with the projector $P_f$
that projects into the subspace spanned by all the blades that have a number that codes the halted machine in any of the
subspaces $4i+2$. If
\be
P_f\psi_3\neq 0
\ee
then this machine will after a certain number of steps halt.

\section{Summary}

There are five properties of GA that are being exploited in this paper.
\

1. Parallelism. Each number that appears in the computation is coded as a blade in a certain subspace. The elements of the
subspace are multivectors which can be linear combinations of some or even all possible blades. This is the analogue of quantum
systems' property of being in a superposition of states. Along with the linearity of the operators this allows simultaneous
computation of an arbitrary number of instances of an operator. This property does not appear in classical computation.
\

2. Transparency. We assume that there is no Heisenberg's Uncertainty Principle and we can just look at the multivector at
any time and extract any data necessary with ease. This property does not appear in quantum computation.
\

3. Arbitrary operations. There are no laws that forbid us to perform nonunitary operations. This property is also absent
in quantum computation. Three first properties were exploited in sections II and III to show two examples of superefficient
algorithms.
\

4. Infinite size of space. The space for GA can be infinite. For computation it means the possibility of having an infinite
memory or computing an infinite number of steps as done in section IV. Classical computers lack both of these properties.
One could argue that a quantum system with an infinite number of degrees of freedom could play the role of an infinite memory,
but even if we accept this reasoning then the ability to run the infinite number of iterations which would mean either
running for an infinite time or having an infinite copies of the system prohibiting quantum computation from performing
similar algorithms.
\

5. Operator/state duality. The fact that there is no distinction between the state and the operator (both are multivectors
in the same space), does not allow by itself any particular operations forbidden in quantum and classical computation
but carries a certain elegance that should always be considered during the choice of representation.
\

Finally, some remarks on the multivector $PU\psi_0$ from section IV. It is the whole solution to the Halting Problem since
the subsequent steps are only needed to retrieve the outcome already found. The lack of the contradiction between Turing's solution
of Hilbert's problem comes from the fact that every definition of an algorithm puts a stress on the finite number of steps
needed to perform it. Here the number of steps is infinite and any quantum or classical computer would need an infinite
time to construct that multivector, but due to properties of GA it is possible to translate this infinite time into the infinite
space which is not uncommon in computation.
\

Though it is not my aim to find physical systems that can perform this ''pseudoalgorithm'', I would like to draw the attention
to the point of view advocated by Penrose \cite{P1,P2}, that the world at the basic level is governed by an evolution which
is not computable. If this is really the case, the description of the world by Clifford algebras, which is becoming increasingly
popular among physicists and, as shown here, can be used to solve noncomputable problems, may be the right choice.

\section{Acknowledgments}
I would like to thank Marek Czachor for many hours of enlightening discussions.

\end{document}